# Comparison Between, and Validation Against an Experiment of, a Slowly-Varying Envelope Approximation Code and a Particle-in-Cell Simulation Code for Free-Electron Lasers

L.T. Campbell,[1,2] H.P. Freund,[3] J.R. Henderson,[1,2] B.W.J. McNeil, [1,4] P. Traczykowski[1,2,4] and P.J.M. van der Slot[5]
[1]University of Strathclyde (SUPA), Glasgow G4 0NG, United Kingdom
[2]ASTeC, STFC Daresbury Laboratory and Cockcroft Institute, Warrington WA4 4AD, United Kingdom
[3]University of New Mexico, Albuquerque, New Mexico, 87131 United States of America
[4]Cockcroft Institute, Warrington WA4 4 AD, United Kingdom
[5]Mesa[+] Institute for Nanotechnology, University of Twente, Enschede, the Netherlands

Free-electron lasers (FELs) operate at wavelengths down to hard x-rays, and are either seeded or start from noise. There is increasing interest in x-ray FELs that rely on Self-Amplified Spontaneous Emission (SASE), and this involves increasing simulation activity in the design, optimization, and characterization of these x-ray FELs. Most of the simulation codes in use rely on the Slowly-Varying Envelope Approximation (SVEA) in which Maxwell's equations are averaged over the fast time scale resulting in relatively small computational requirements. While the SVEA codes are generally successful, the predictions of these codes sometimes differ in various aspects of the FEL interaction. In contrast, Particle-in-Cell (PiC) simulation codes do not average Maxwell's equations and are considered to be a more complete model of the underlying physics. Unfortunately, they require much longer run times than SVEA codes and have not been validated by comparison with experiment as often as the SVEA codes. In order to remedy this, and to resolve issues that arise due to different predictions between the SVEA codes, we present a comparison between one SVEA code (MINERVA) and a PiC simulation code (PUFFIN) with the experimental measurements obtained at the SPARC SASE FEL experiment at ENEA Frascati. The results show good agreement between the two codes and between the codes and the experiment. Since the formulations of the two codes share no common elements, this validates both formulations and demonstrates the capability to model the FEL interaction from the start of the undulator through the undulator and into deep saturation.

## I. INTRODUCTION

While free-electron lasers (FELs) have been intensively studied since the 1970s, new developments and concepts keep the field fresh. Intensive work is ongoing into new FEL-based light sources that probe ever shorter wavelengths with a variety of configurations. There presently exists a large variety of FELs ranging from long-wavelength oscillators using partial wave guiding to ultraviolet and hard x-ray FELs that are either seeded or starting from noise (*i.e*., Self-Amplified Spontaneous Emission or SASE). As new FEL light sources come on-line, interest will grow in shorter pulses, new spectral ranges and higher photon fluxes. The increasing activity in the design and construction of FEL light sources is associated with increasing simulation activity to design, optimize, and characterize these FELs.

Most of the FEL simulation codes in use at the present time can be categorized as either Slowly-Varying Envelope Analyses (SVEA) or Particle-in-Cell (PiC) simulations. In the SVEA, the optical field is represented by a slowly-varying amplitude and phase in addition to a rapid sinusoidal oscillation. The field equations are then averaged over the rapid sinusoidal time scale and, thereby, reduced to equations describing the evolution of the slowly-varying amplitude and phase. Within the context of the SVEA, FEL simulation codes fall into two main categories where the particle trajectories are found by first averaging the trajectories over an undulator period (the so-called wiggler-averaged-orbit approximation), or by the direct integration of the Newton-Lorentz equations. There is a further distinction between the SVEA codes based upon the optical field representation, and codes have been written using either a grid-based field solver or a superposition of optical modes. Simulation codes using the wiggler-averaged-orbit analysis in conjunction with a grid-based field solver include (but are not limited to) GINGER [1], GENESIS [2], and FAST [3]. In contrast, SVEA codes that integrate the Newton-Lorentz equations in conjunction with a Gaussian mode superposition for the optical fields include MEDUSA [4] and MINERVA [5]. One common feature of all the SVEA codes, however, is the way in which time-dependence is treated. The fast time scale average results in a breakdown of the optical pulse into temporal *slices* each of which is one wave period in duration. The optical slices slip ahead of the electron slices at the rate of one wavelength per undulator period. As a result, the SVEA codes integrate each electron and optical slice from $z \rightarrow z + \Delta z$ and then allow the optical slice to slip ahead of the electron slices. These codes have been extremely successful in modeling FELs; however, their predictions are not always identical for all aspects of the FEL interaction.

In contrast, PiC codes do not average Maxwell's equations and are considered to represent a more fundamental model of the physics of FELs. A PiC code makes no average over the rapid sinusoidal oscillation and integrates the Newton-Lorentz equations for the particles as well as Maxwell's equations for the fields. As a result, PIC codes require substantially more computational resources than SVEA codes and are not so commonly in use and have not been as extensively validated against experiments as have the SVEA codes. At the present time, the primary PiC code for FEL simulations is PUFFIN [6].

In view of this, we undertake in this paper to present a comparison of one SVEA code (MINERVA) and a PiC code (PUFFIN) with experimental measurements. The properties/capabilities of these codes have been presented in the literature and will not be discussed here other than to emphasize that while MINERVA applies the SVEA it does not average the Newton-Lorentz equations over the undulator period. As such, both PUFFIN and MINERVA integrate the particle trajectories in the full magnetostatic and electromagnetic field representations. Other than this, the two codes share no common elements. In particular, the particle loading algorithms used to treat start-up from noise are different. MINERVA uses an adaptation of the algorithm described by Fawley [8] while PUFFIN uses an algorithm developed by McNeil et al. [9].

The organization of the paper is as follows. Our purpose in this paper is to compare the simulation results obtained by the two codes and to "validate" the codes by comparison with experimental measurements taken in a SASE FEL. To this end, comparisons between PUFFIN and MINERVA and between the two codes and experimental measurements at the "Sorgente Pulsata ed Amplificata di Radiazione Coerente" (SPARC) experiment which is a SASE FEL located at ENEA Frascati [7] are presented in Section II. A summary with conclusions is given in Section III.

**II. THE SPARC SASE FEL**

The "Sorgente Pulsata ed Amplificata di Radiazione Coerente" (SPARC) experiment is a SASE FEL located at ENEA Frascati [7]. The best estimate for the experimental parameters of SPARC are summarized in Table 1 and are as follows. The electron beam energy was 151.9 MeV, with a bunch charge of 450 pC, and a bunch width of 12.67 psec. A parabolic temporal bunch profile was used in MINERVA while PUFFIN employed a Gaussian temporal profile. In practice, these two choices for the temporal profiles did not result in any significant differences in the predictions found by the two codes or between the codes and the experimental measurements. The $x$ and $y$ emittances were 2.5 mm-mrad and 2.9 mm-mrad respectively, and the rms energy spread was 0.02%. There were six undulators in the experiment, each of which was 77 periods in length (with one period for the entrance up-taper and another for the exit down-taper) with a period of 2.8 cm and an amplitude of 7.88 kG.

The experiment employed six undulators for an overall length of about 15 meters, but this was too short to reach saturation given the bunch charge. In order to compare the codes in the saturated regime, therefore, we extended the undulator/FODO lattice to include 11 undulators out to a distance of about 28 meters. As a result, the experimental data is used to anchor the validation of the codes in the start-up and exponential growth regions, while the code results are compared for the initial start-up, exponential growth and deep saturation regimes. The gaps between the undulators were 0.4 m in length and the quadrupoles (0.053 m in length with a field gradient of 0.9 kG/cm) formed a strong focusing lattice and were located 0.105 m downstream from the exit of the previous undulator. Note that the quadrupole orientations were fixed and did not alternate. The electron beam was matched into the undulator/focusing lattice. The resonance occurred at a wavelength of 491.5 nm. In the experiment, the pulse energies were measured in the gaps after each undulator segments by opening the gaps in successive undulators, thereby detuning the FEL interaction, in the further downstream undulators [7].

| Electron Beam | |
|---|---|
| Energy | 151.9 MeV |
| Bunch Charge | 450 pC |
| Bunch Duration | 12.67 psec |
| $x$-Emittance | 2.5 mm-mrad |
| $y$-Emittance | 2.9 mm-mrad |
| rms Energy Spread | 0.02% |
| rms Size ($x$) | 132 microns |
| $\alpha_x$ | 0.938 |
| rms Size ($y$) | 75 microns |
| $\alpha_y$ | -0.705 |
| **Undulators** | |
| Period | 2.8 cm |
| Length | 77 Periods |
| Amplitude | 7.8796 kG |
| $K_{rms}$ | 1.457 |
| Gap Length | 0.40 m |
| **Quadrupoles** | |
| Length | 5.3 cm |
| Field Gradient | 0.9 kG/cm |

Table 1: Parameters of the SPARC FEL experiment.

The simulated propagation of the beam through the undulator/quadrupole lattice is shown in Fig. 1, where we plot the beam envelope in $x$ (blue, left axis) and $y$ (red, right axis) versus position as determined by MINERVA. The PUFFIN propagation results were similar. Observe that the beam is well-confined over the 28 meters of the extended lattice with an average beam size of approximately 115 microns.

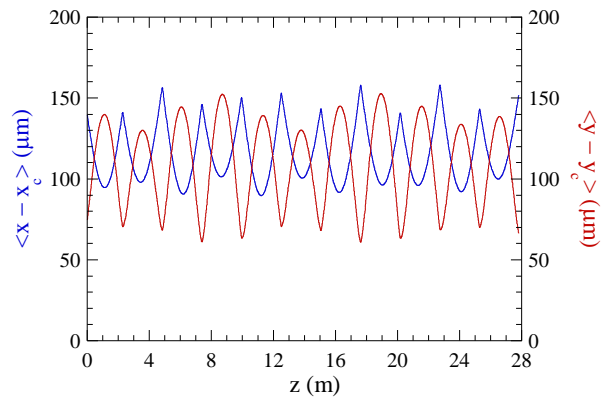

Fig. 1: MINERVA simulation of the beam propagation through the undulator/quadrupole lattice.

A comparison of the evolution of the pulse energy as found in MINERVA and PUFFIN, and as measured in the experiment, is shown in Fig. 2 where the MINERVA simulation is indicated by the blue line and the PUFFIN simulation is indicated by the green line. The measured pulse energies for a sequence of shots are indicated by the red markers where the error bars indicate the standard deviation over a sequence of shots (data courtesy of L. Giannessi). Observe that the agreement between the two codes, and between the codes and the measured pulse energies, are excellent over the entire range of the experiment.

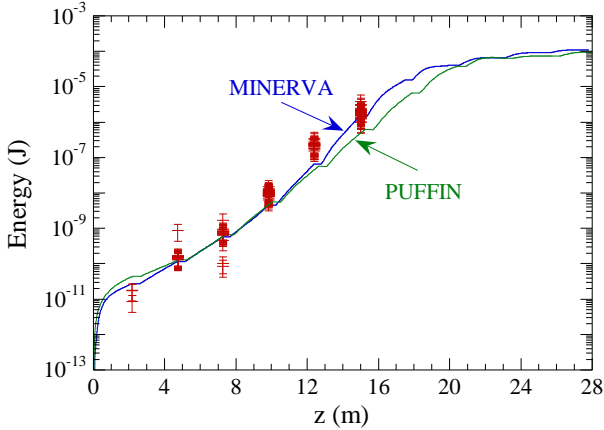

Fig. 2: Comparison of simulation results with PUFFIN and MINERVA and the measured pulse energies versus distance through the undulator (data courtesy of L. Giannessi).

We remark that the exponential growth region starts in the second undulator and that the start-up region is encompassed in the first undulator segment. The experimental measurements indicate that the pulse energy after the first undulator falls into the range of about $8.4 \times 10^{-12}$ through $1.74 \times 10^{-11}$ J while MINERVA yields a pulse energy of $2.52 \times 10^{-11}$ J and PUFFIN yields $4.02 \times 10^{-11}$ J. That the experimental value is somewhat lower than the simulations is to be expected as it is measured at some distance downstream from the first undulator segment while the codes evaluate the pulse energy at every location along the undulator line. The greatest differences between the simulations are found in the sixth and seventh undulators; however, the differences between the codes are smaller than the uncertainty in the experimental measurements. Hence, the simulation results are in relatively close agreement with the experiment and with each other. This agreement is an important observation since the particle loading algorithms in the two codes share no commonality. Apart from differences that might derive from the parabolic versus Gaussian temporal profiles and the different particle loading algorithms, another source of the difference in the slightly higher start-up noise in PUFFIN is the fact that PUFFIN naturally includes a wider initial spectral range than MINERVA.

The exponential growth region starts in the second undulator and the two codes are in close agreement with each other and with the experimental measurements out to the end of the sixth undulator. These results are in substantial agreement with the parameterization developed by Ming Xie [10]. Using a $\beta$-function of about 2 m, we find that the Pierce parameter $\rho \approx 2.88 \times 10^{-3}$ and that this parameterization predicts a gain length of 0.67 m, and a saturation distance of 18.1 m (including the additional 3.2 m represented by the gaps between the undulators). This is in reasonable agreement with the simulations which indicate that saturation occurs after between about 18 – 20 meters of undulator/FODO line.

Finally, the predictions of the two codes in the saturation regime after about 20 m are also in remarkable agreement. After 28 m of undulator/FODO lattice Puffin predicts a pulse energy of 96 μJ while MINERVA predicts 111 μJ which constitutes a difference of about 13.5 %.

The larger initial spectral linewidth excited in the start-up region exhibited by PUFFIN is shown more clearly in Fig. 3 which presents a comparison between the evolution of the relative linewidth as determined from PUFFIN and MINERVA and by measurement (data courtesy of L. Giannessi). It is clear that PUFFIN predicts a wider initial spectrum than MINERVA. This is consistent with the full dynamics modelled by a PiC code such as PUFFIN which includes the generation of the wider bandwidth incoherent spontaneous emission. Exponential gain due to the resonant FEL interaction starts in the second undulator and this is expected to rapidly overcome any incoherent synchrotron radiation from the start-up region in the first undulator. In view of this, the PUFFIN results converge rapidly to that found by MINERVA and to the measured linewidths after the second undulator. Note, however, that the measured linewidth after the first undulator seems to be in better agreement with the MINERVA result, but this may be due to the bandwidth of the detector. Agreement between the simulations and the measured linewidth is within about 35% after 15 m. As shown in the figure, the predicted linewidths are in substantial agreement with the experimental measurements, and good agreement between the codes is found over the entire range of integration through the saturated regime.

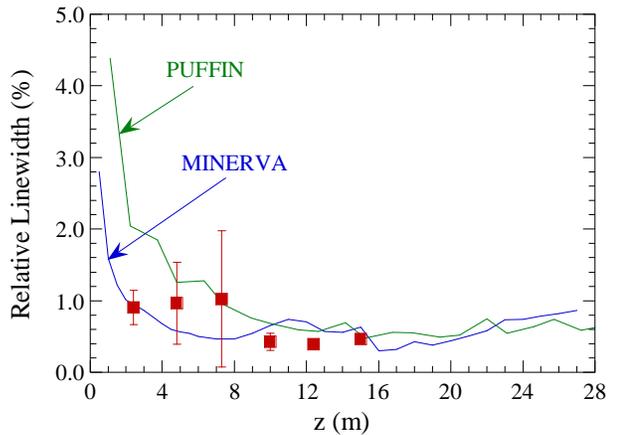

Fig. 3: Comparison of the measured relative linewidth in red (data courtesy of L. Giannessi) with that found in the simulations (blue for MINERVA and green for PUFFIN).

## III. SUMMARY AND CONCLUSION

In this paper, we have described a comparison between FEL simulation codes based on the SVEA formulation (MINERVA) and on a PiC formulation (PUFFIN). The two codes have been run in simulation of the SPARC SASE FEL at ENEA Frascati. Good agreement has been found both between the two codes and between the codes and the experiment, thereby validating both formulations.

This is significant because these two formulations have virtually no elements in common, and we can conclude from this that they both faithfully describe the physics underlying FELs. In particular, the agreement between the codes and the experimental measurements regarding the start-up regime in the SPARC FEL validates the different particle loading algorithms in both codes.

One limitation of the SVEA models derives from the fast time scale average which means that these codes cannot treat ultra-short pulse production. This limitation is not present in PiC codes; hence, the validation of PUFFIN implies that it may be a useful model for future ultra-short pulse propagation experiments.

## ACKNOWLEDGEMENTS

The research used resources of the Argonne Leadership Computing Facility, which is a DOE Office of Science User Facility supported under contract DE-AC02-06CH11357. We also thank the University of New Mexico Center for Advanced Research Computing, supported in part by the National Science Foundation, for providing high performance computing resources used for this work.

Funding is also acknowledged via the following grants: Science and Technology Facilities Council (Agreement Number 4163192 Release #3); ARCHIE-WeSt HPC, EPSRC grant EP/K000586/1; John von Neumann Institute for Computing (NIC) on JUROPA at Jülich Supercomputing Centre (JSC), project HHH20.

The authors acknowledge helpful discussions with L. Giannessi.